# Development and Usability Study of Older Adults in Motion-Captured Serious Game Incorporating Olfactory Stimulations

*Joyce S.Y. Lau[1], Zihui Jing[1], Clement P.L. Chan[1], Louis C.F. Ng[1], Wing Chin Kam[2], Kwan Yin Lam[2], Ho Wui Cheung[2], Ho Lam Lau[2] and Junpei Zhong[3]*

*Junpei Zhong: joni.zhong@ieee.org*

[1]Pathfinder Technology Limited, Hong Kong

[2]Faculty of Science and Technology, UOW College Hong Kong, Hong Kong

[3]Department of Digital Innovation and Technology, Technological and Higher Education Institute of Hong Kong, Hong Kong

## Abstract

**Background and Objectives:**

This study developed SENSO™, a motion-captured virtual reality serious game incorporating multisensory stimulations (visual, auditory, olfactory) to enhance cognitive and motor function in older adults. The objective was to evaluate its usability and performance among healthy seniors, establishing baselines for MCI or dementia risk prediction.

Methods:
SENSO™ features three teahouse-themed games: Dim Sum (selection/placement), Steamer (timing/sequencing), and Cashier (counting/transactions), delivered via motion capture and olfactory cues in a controlled setting. Complete datasets were obtained from 41 older adults (aged 60+). Quantitative analysis included System Usability Scale (SUS) surveys and age-stratified performance metrics (accuracy, completion time) derived from game logs to investigate the user experience.

**Results:**

Participants were grouped by age: 60–69 (n=22), 70–79 (n=17), and ≥80 years (n=2). SUS scores indicated high confidence in system use (mean=82/100), reflecting intuitive design adaptable to diverse technical proficiencies. Performance analysis showed no significant age differences in the Dim Sum task (age-neutral), but pronounced declines in the Steamer task due to higher cognitive/motor demands, and notable trends toward decline in the Cashier task. These findings suggest prioritizing Steamer task adaptations (e.g., simplified interfaces, olfactory cues) to mitigate age-related gaps and enhance therapeutic value for MCI prevention.

**Conclusion:**

SENSO™ demonstrates strong usability and age-differentiated performance among healthy older adults, providing normative baselines. Olfactory-enhanced adaptations for demanding tasks can optimize therapeutic efficacy, supporting non-pharmacological interventions to preserve cognitive function. Future longitudinal studies will validate predictive utility.

# 1. Introduction

## 1.1 Background Information on MCI

Mild Cognitive Impairment (MCI) is a clinical state characterized by a noticeable decline in cognitive abilities, such as memory and thinking skills, that exceeds the expected decline associated with aging but does not significantly interfere with daily activities (Anderson, 2019). MCI is often considered a precursor to dementia, a profound global health challenge affecting approximately 55 million people worldwide (WHO, 2023). Given the lack of effective pharmacological treatments for MCI, non-pharmacological strategies have become a critical focus. While conventional cognitive training has shown benefits, its repetitive nature often leads to low engagement and adherence. Consequently, innovative, engaging alternatives, such as lifestyle modifications and multisensory stimulation, are being actively explored for managing symptoms and potentially slowing disease progression (Gómez-Soria et al., 2022).

Virtual Reality (VR) has emerged as a powerful platform for delivering such stimulation due to its unique features (Brugada-Ramentol et al., 2022; Riaz et al., 2021; Muñoz et al., 2021). VR is an immersive technology that uses 3D computer-generated environments to simulate realistic daily-life scenarios, allowing users to feel a sense of "being there" (Slater, 2018) and interact with dynamic environments in real-time (Mancuso et al., 2020). This ability to recreate lifelike, ecologically valid contexts in a safe and controlled setting (Moreno et al., 2019; Bohil et al., 2011) has made VR a valuable tool in clinical rehabilitation for fostering engagement and presence.

Although the current VR systems mostly rely on the visual modality, olfactory stimulation also holds particular promise due to the olfactory system's direct and unique connection to brain regions vital for memory and emotion, such as the amygdala and hippocampus (Brennan et al., 1990; Soudry et al., 2011). The integration of olfactory cues with VR can significantly enhance immersion, memory recall, and therapeutic outcomes, yet studies rigorously combining these two modalities are limited. Furthermore, while most VR interventions focus solely on cognitive training, dual-task interventions that combine cognitive and physical challenges are critical for addressing the functional decline in older adults. Technical developments such as motion-captured devices also address the need (Rucco et al., 2016; Bishnoi & Hernandez, 2020). The integration of multisensory has been utilised for long as a therapeutic approach, which engages multiple senses simultaneously. It has gained increasing attention for its potential to enhance cognitive, emotional, and physical well-being in individuals with cognitive impairments (Sánchez et al., 2012; Manippa et al., 2022; Yang et al., 2021; Helbling et al., 2023). But our study aims at developing a novel game which integrates all these technologies in one system to address the MCI cognitive training needs.

## 1.1 Study Aim and Novel Contribution

Drawing upon these converging lines of evidence, this paper presents the development and usability study of a novel intervention: a motion-captured serious game incorporated with olfactory stimulations, titled SENSO™ , designed specifically for healthy older adults (as a precursor to testing in MCI populations). The game simulates everyday scenarios, such as select and pick up dim-sum and money calculation, using markerless motion capture for intuitive, dual-task interactions and synchronized scents to boost engagement and memory association.

The primary objectives of this pilot study are to:

1. Describe the game's design and technical implementation, specifically the integration of markerless motion capture and synchronized olfactory cues.
2. Evaluate the usability, system acceptance, and user experience of the integrated system in a cohort of healthy older participants, using the System Usability Scale (SUS) and questionnaires.
3. Explore preliminary quantitative performance metrics (e.g., task accuracy and completion time) within the serious game.

By combining non-immersive VR, markerless motion capture, and olfactory stimulation into a user-centered serious game, this work addresses a significant gap in multisensory, dual-task interventions, contributing to innovative, highly engaging approaches for cognitive and physical health in aging.

---

# 2. Literature Review

## 2.1 Multisensory Stimulation and Virtual Reality in Cognitive Interventions

Multisensory stimulation, the strategic integration of sensory inputs (sight, hearing, touch, smell, and taste), forms the basis of therapeutic experiences designed to enhance cognitive function (Helbling et al., 2023). VR provides a highly effective medium for delivering this stimulation by creating controlled, yet realistic, environments that engage users in meaningful, everyday activities (Mancuso et al., 2020).

Studies have shown a positive impact of VR-based interventions, including serious games, on cognitive domains in individuals with cognitive impairment, with improvements observed in memory, attention, and executive function (Kim et al., 2019; Optale et al., 2009). The therapeutic mechanism is thought to involve the recruitment of neuroplastic mechanisms through synchronous, rich sensory inputs (Mancuso et al., 2020; Zhu et al., 2021). A recent systematic review of VR-based multisensory training in MCI reported significant improvements in global cognition (measured by MoCA and MMSE), underscoring VR's capacity for sensory integration and neural network recruitment (Khotbehsara et al., 2025). Furthermore, preliminary experimental studies suggest that adding non-visual cues (e.g., tactile and olfactory) to immersive VR enhances memory retrieval and spatial judgment compared to visual-only formats (Lee and Xin-ting., 2024).

## 2.2 Olfactory Stimulation in MCI and Dementia Care

Olfactory stimulation offers a powerful, non-pharmacological pathway to the brain, providing direct access to memory and emotional centers (Marin et al., 2018). The olfactory system's relative preservation in the early stages of dementia makes it a valuable tool for triggering implicit memories.

Growing evidence supports the neuroplastic effects of olfaction. For instance, minimal olfactory enrichment, even during sleep, has been shown to lead to significant improvements in cognitive functioning in older adults (Woo et al., 2023). Moreover, olfactory training can increase grey matter volume in critical regions, including the hippocampus and entorhinal cortex (Aïn et al., 2019; Rezaeyan et al., 2021).

The therapeutic application of olfaction is particularly relevant for reminiscence therapy, which is a psychosocial intervention method leveraging familiar cues to evoke autobiographical memory (Cotelli et al., 2012; Lau et al., 2024). Familiar scents, such as culturally relevant food odors, can evoke past experiences, helping individuals with cognitive decline reconnect with their personal histories (Malloggi et al., 2021). The critical factor for efficacy is odor familiarity, which has been identified as the single strongest predictor of pleasantness ratings, surpassing the inherent chemical properties of the odorants themselves (Arshamian et al., 2022; Sorokowski et al., 2024). This highlights the necessity of selecting culturally relevant or personally familiar odors in therapeutic design. While multisensory environments (MSEs) have demonstrated feasibility in reducing agitation and anxiety in dementia patients (Riley-Doucet & Dunn, 2013), the integration of olfaction into fully immersive VR platforms remains an under-explored, yet highly promising, area.

## 2.3 Motion Captured VR for Dual-Task Training in Older Adults

Age-related decline in dual-task performance—the ability to perform a cognitive task simultaneously with a physical task—is a critical predictor of fall risk in older adults (Bishnoi & Hernandez, 2020). Motion-captured Virtual Reality (VR) dual-task training has emerged as a promising rehabilitation modality to mitigate these risks (Rucco et al., 2016).

Motion capture, which can range from non-immersive markerless systems (e.g., Microsoft Kinect) to immersive Head-Mounted Displays (HMDs) with wearable trackers, provides real-time movement feedback and enables natural user interaction (Lam et al., 2023; Honzíková., et al., 2025). Interventions are designed to foster "cognitive-motor synergy" by requiring users to perform tasks like mathematical calculation or spatial memory while simultaneously executing physical movements, such as reaching or stepping (Hassandra et al., 2025; Omon et al., 2019). This gamified combination mimics real-life complexity, training the brain to allocate resources more efficiently and reducing the "dual-task cost."

Meta-analytic evidence supports the clinical efficacy of these interventions, showing significant improvements in both spatiotemporal gait parameters and executive functions compared to traditional single-task training (Wei et al., 2025; Wenk et al., 2022). Furthermore, the engaging nature of gamified platforms often leads to high adherence rates (Aldardour, & Alnammaneh, 2025). The use of markerless motion capture is particularly relevant for older adults, as it eliminates the need for cumbersome wearables, promoting ease of use and accessibility.

## 2.4 Serious Games for MCI

Serious games are digital games with a primary therapeutic, educational, or training purpose (Manca et al., 2020). They leverage gamification elements (e.g., challenges, rewards, and progression) to improve motivation and therapeutic outcomes. For MCI, these games are specifically engineered to engage multiple cognitive domains—memory, attention, and executive function—within an interactive and motivating framework (Yun et al., 2020).

Systematic reviews and meta-analyses consistently report that serious games, especially those integrating physical activity ("exergames"), yield statistically significant improvements in global cognition and executive functions (Abd-Alrazaq et al., 2022; Zhao et al., 2020). The high efficacy of VR-based games (Hedges's $g = 0.60$ for cognitive rehabilitation) is attributed to their ability to create immersive, multisensory environments that enhance "cognitive-motor synergy" (Tortora et al., 2024).

A key factor in efficacy is the ecological validity of the game design. Games that simulate Activities of Daily Living (ADLs)—such as virtual supermarket shopping or preparing a meal—are highly effective because they train functional skills with direct real-world transferability. Furthermore, the use of user-centered design, where difficulty adapts dynamically to the user's performance, is essential for maintaining engagement and preventing frustration (Seyderhelm, A. J., & Blackmore, 2021; Schaumburg et al., 2025).

## 2.5 Identified Gap and Study Rationales

Prior research has established the individual therapeutic value of multisensory VR, motion-captured dual-task training, and olfactory stimulation in cognitive care. However, a significant gap remains: there is a lack of cohesive studies that fully integrate all three elements into a single, ecologically valid intervention for older adults.

Specifically, current VR serious games often neglect the olfactory sense, limiting the potential for deeply embedded memory and emotional engagement. Furthermore, interventions that combine olfaction with VR rarely incorporate real-life and motion-captured interaction, thus missing the benefits of crucial dual-task cognitive-motor training.

To address this, the SENSO™ serious game was developed. It uniquely combines:

1. **Non-Immersive VR-Based Serious Game:** For high engagement and training of IADL-relevant cognitive skills (e.g., calculation and object selection).
2. **Markerless Motion Capture:** For natural, hand gestural, dual-task physical and cognitive engagement.
3. **Synchronized Olfactory Stimulation:** To enhance immersion, memory encoding, and emotional resonance using culturally relevant scents.

This study aims to rigorously evaluate the user experience and usability of this novel, integrated system in a cohort of healthy older adults, providing the necessary foundation for future efficacy trials in the MCI population.

## 3. Methods

### 3.1 SENSO™ System - A Multisensory Game Therapy

### 3.1.1 Development of the SENSO™ System Based on the MCI-Game Therapy Experience (MCI-GaTE) Serious Game Framework

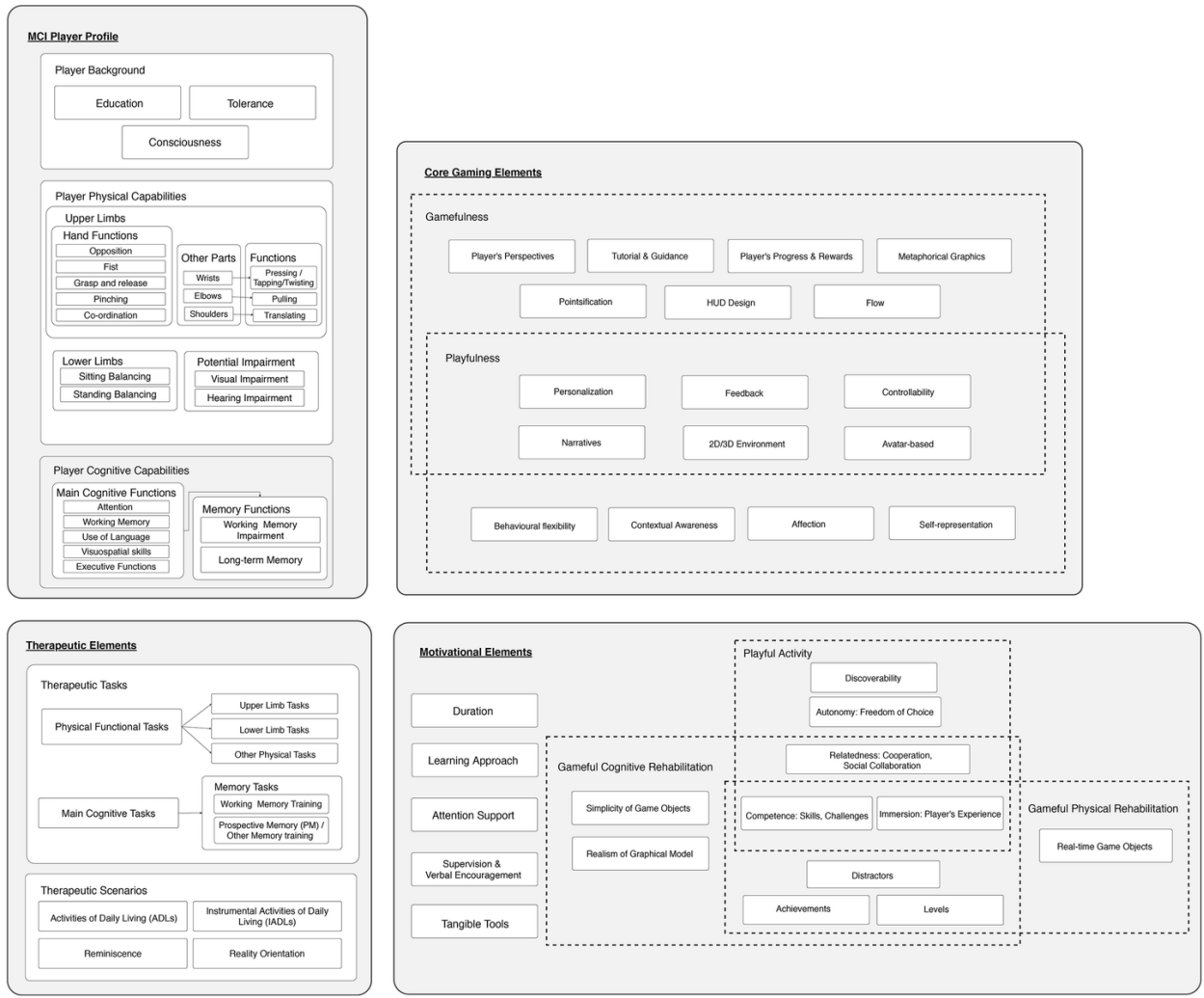

*Figure 1. The MCI-Game Therapy Experience (MCI-GaTE) framework: A serious game framework designed for individuals with mild cognitive impairment. (Lau & Agius, 2021)*

The development of the SENSO™ system was guided by the MCI-Game Therapy Experience (MCI-GaTE) framework, developed by Lau and Agius (2021) and illustrated in Figure 1, which serves as a structured design model for creating serious games tailored to individuals with mild cognitive impairment (MCI). The framework emphasizes four interrelated themes: (1) MCI Player Profile, (2) Therapeutic Elements, (3) Core

Gaming Elements, and (4) Motivational Elements. Each providing essential design considerations to ensure therapeutic effectiveness, usability, and engagement among older adults.

The MCI Player Profile served as a foundational reference for user-centered design. It defines the cognitive, physical, and sensory characteristics of adults with MCI, including possible limitations in memory, attention, and motor coordination. Based on these insights, the SENSO™ interface was simplified with clear visual prompts, minimal textual instructions, and slower-paced interactions. Input mechanisms were tailored to accommodate reduced fine motor control, using large, motion-capture–based gestures instead of complex controller inputs. The inclusion of olfactory cues further supported multisensory engagement and memory associations, aligning with the sensory processing needs of the MCI population.

The Therapeutic Elements of the MCI-GaTE framework guided the integration of evidence-based cognitive and motor training components into the SENSO™ game design. Each of the three core games, Dim Sum, Steamer, and Cashier, was developed around instrumental daily living activities that stimulate memory, attention, and executive functions, while simultaneously promoting upper limb movement and bilateral coordination. The tasks were structured with adaptive difficulty levels and automated feedback, enabling participants to experience a sense of mastery while remaining cognitively challenged.

The Core Gaming Elements defined how gameplay mechanics, interface flow, and feedback systems were implemented to create an intuitive and engaging experience. The SENSO™ environment emphasized ecological validity through realistic game scenarios and goal-oriented tasks, reflecting meaningful real-world activities rather than abstract puzzles. The motion-capture interface allowed natural interactions such as grasping, translating, or coordinating, which translated directly to in-game actions. Real-time feedback and cumulative scoring reinforced accuracy and sustained focus throughout the game sessions.

Finally, the Motivational Elements were integrated to enhance user enjoyment, persistence, and emotional engagement, which are critical factors for long term therapeutic adherence. The game incorporated immediate performance feedback, positive reinforcement through visual and auditory cues, and progressive task structures that rewarded improvement over time. The inclusion of familiar cultural themes, such as handling dim sums or managing transactions, also contributed to intrinsic motivation by resonating with players' personal and social contexts.

Taken together, these four design dimensions ensured that the SENSO™ system not only aligned with therapeutic objectives but also fostered cognitive engagement, physical activation, and emotional well-being. By operationalizing the principles of the MCI-GaTE framework, SENSO™ represents a balanced integration of rehabilitation science and game design, optimized for accessibility and sustained use among older adults with mild cognitive impairment.

### 3.1.2 SENSO™ System Architecture

The SENSO™ system was implemented in the Unity game engine (United Technologies) on a Windows PC, providing a flexible and robust environment for real-time interaction, data logging, and visualization. Graphical user interfaces (GUIs) and three-dimensional (3D) assets were created using Autodesk Maya and the Adobe Creative Suite to ensure ergonomic, visually clear, and age-appropriate designs tailored to older adults.

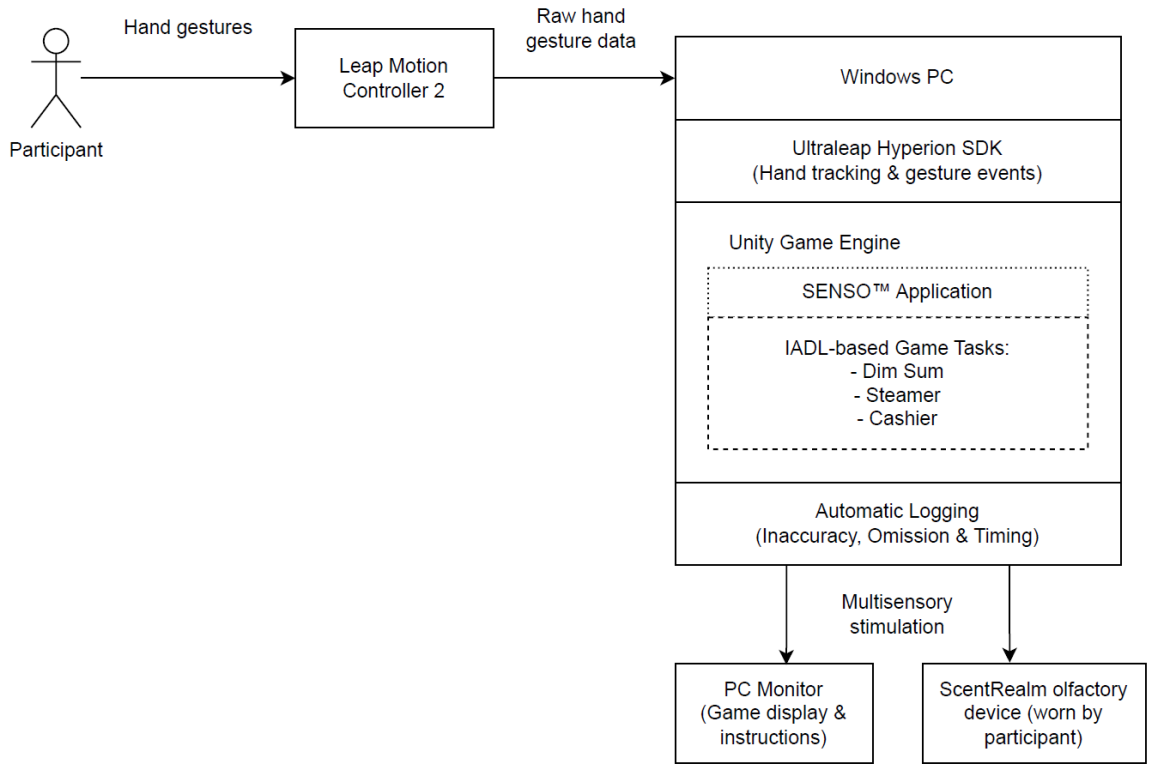

*Figure 2. SENSO<sup>TM</sup> System Architecture*

SENSO™ system architecture is illustrated in Figure 2. A Leap Motion Controller 2 (Ultraleap) is connected to the PC and positioned in front of the participant to capture hand gesture and upper limb motions. Gesture data are processed via the Ultraleap Hyperion SDK and mapped to in-game actions. and the SENSO™ application (www.senso.games). In the application, three IADL-based game modules (Dim Sum, Steamer, and Cashier) are built. Participants face the PC monitor while wearing a ScentRealm™ olfactory device (Scentrealm) for multisensory stimulation. Performance metrics, including inaccuracy, omission, and timing, are automatically logged.

User interaction relies exclusively on markerless motion capture by using Leap Motion System.. This configuration enables precise and responsive hand tracking and gesture recognition, allowing participants to perform natural movements such as grasping, releasing, and sorting without handheld controllers. The motion-based interface enhances accessibility and motor engagement, promoting intuitive interaction even among participants with limited technological familiarity.

The SENSO™ gameplay design is grounded in the principles of Instrumental Activities of Daily Living (IADL), emphasizing simulated three local daily tasks that engage both cognitive and functional skills:

- Dim sum (Figure 3)(working memory and visuomotor coordination).

  Participants first view a subset of dim sum items for memorization (Figure 3a). They then select the correct item from a virtual cart and place it on a designated area on the table (Figure. 3b). This module primarily targets short-term/working memory and hand–eye coordination.

- Steamer (Figure 4)(executive function, timing, and sequencing)

  Participants move selected dim sum items into a steamer, manage steaming times using visual countdown cues (Figure. 4a), and then transfer them to a serving area (Figure. 4b). The olfactory device emits corresponding food scents during steaming and a "burnt" odor if the food is overcooked (Figure 4c). This module targets planning, inhibition, temporal judgment, and response to feedback.

- Cashier (Figure 5)(calculation and motor precision)

  Participants receive a bill total and a payment amount and must return the correct change by grasping virtual banknotes/coins from a cash register and placing them into a holder. This task engages numerical reasoning, working memory, and upper-limb control in a simulated transactional scenario.

Through these goal-oriented scenarios, SENSO™ offers an ecologically valid training environment that supports cognitive and physical domains relevant to independent living.

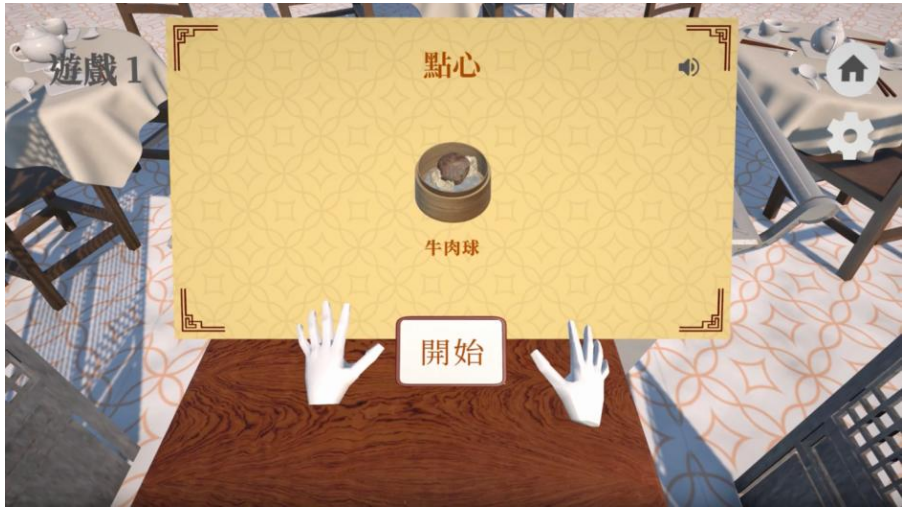

*Figure 3a: Dim Sum game objects displayed for player memorization, engaging working memory.*

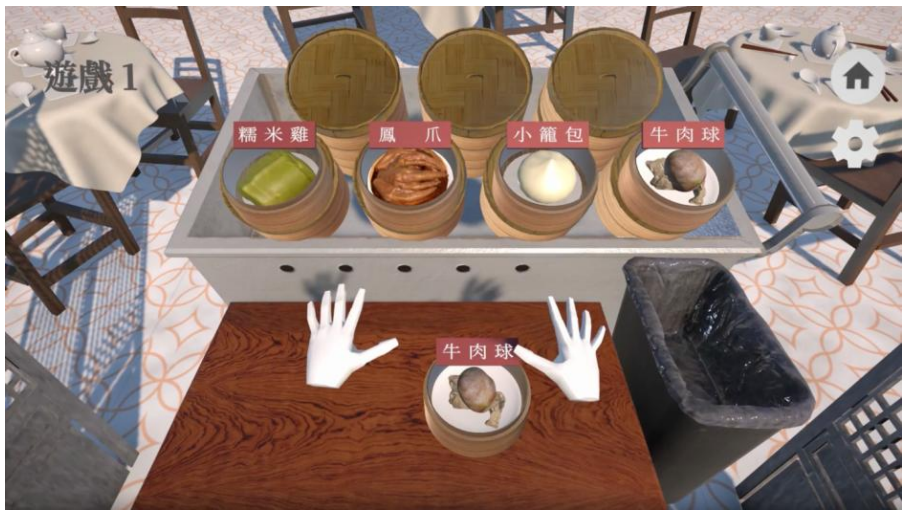

*Figure 3b: Player grasps correct object from dim sum cart and places it on table.*

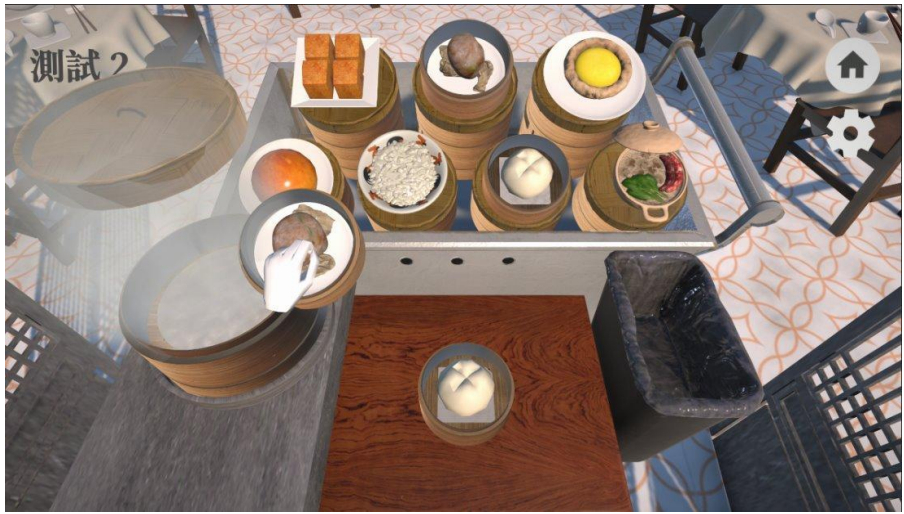

*Figure 4a: The player is required to grasp the correct dim sum and put it into the steamer.*

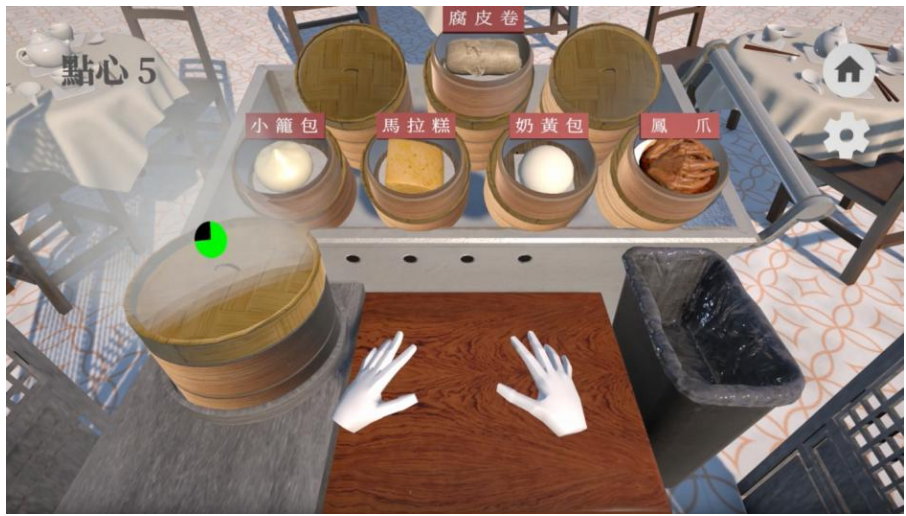

*Figure 4b: The steamer timer will turn green as a visual cue to notify the player to remove the steamer lid.*

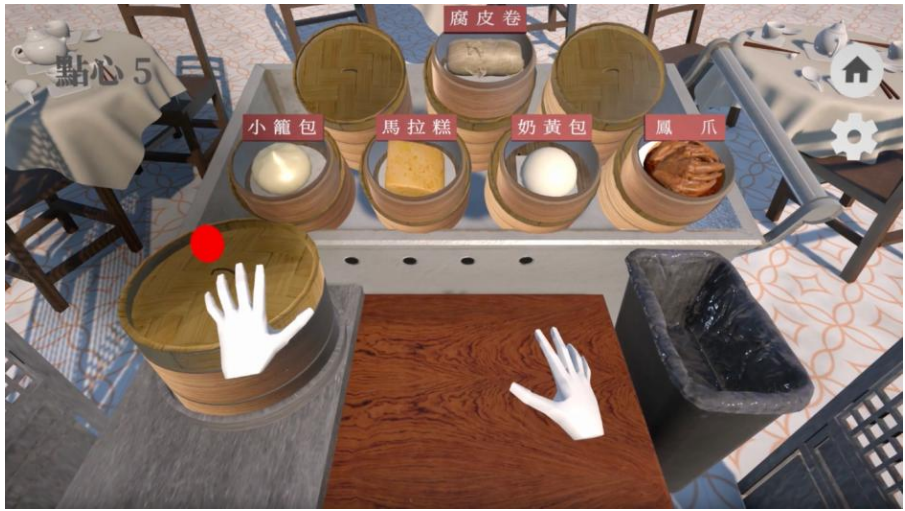

*Figure 4c: The steamer timer will turn red as a visual cue to alert the player that the food is*

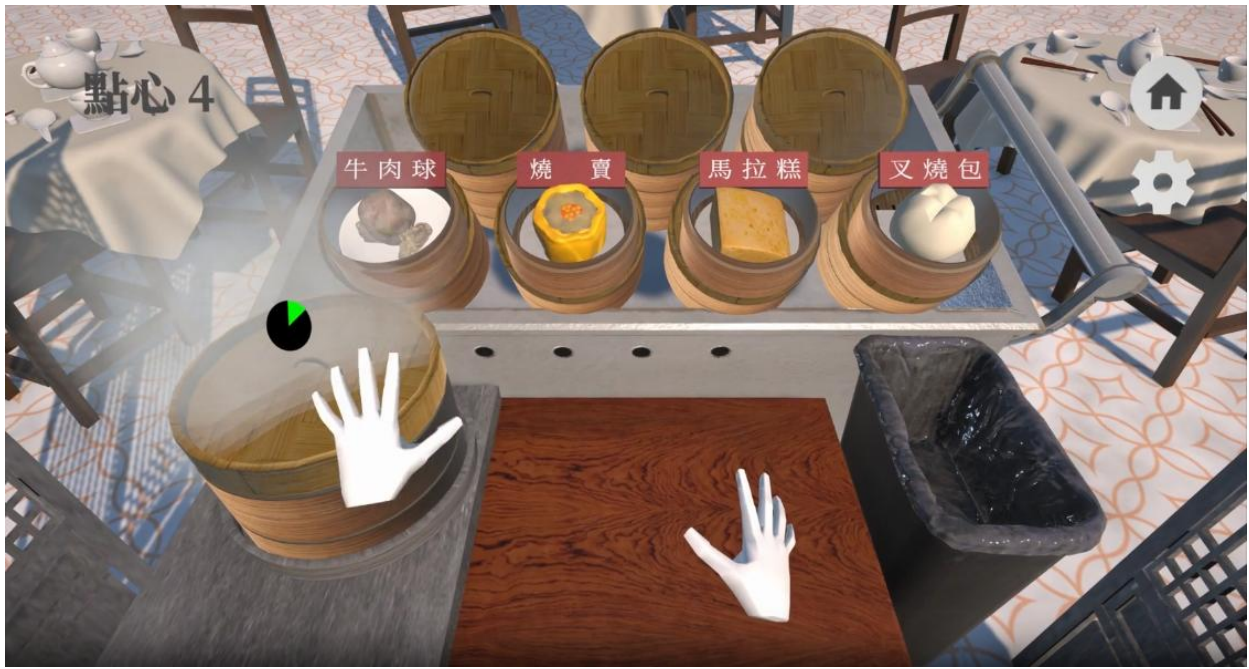

*Figure 4. Steamer game scene*

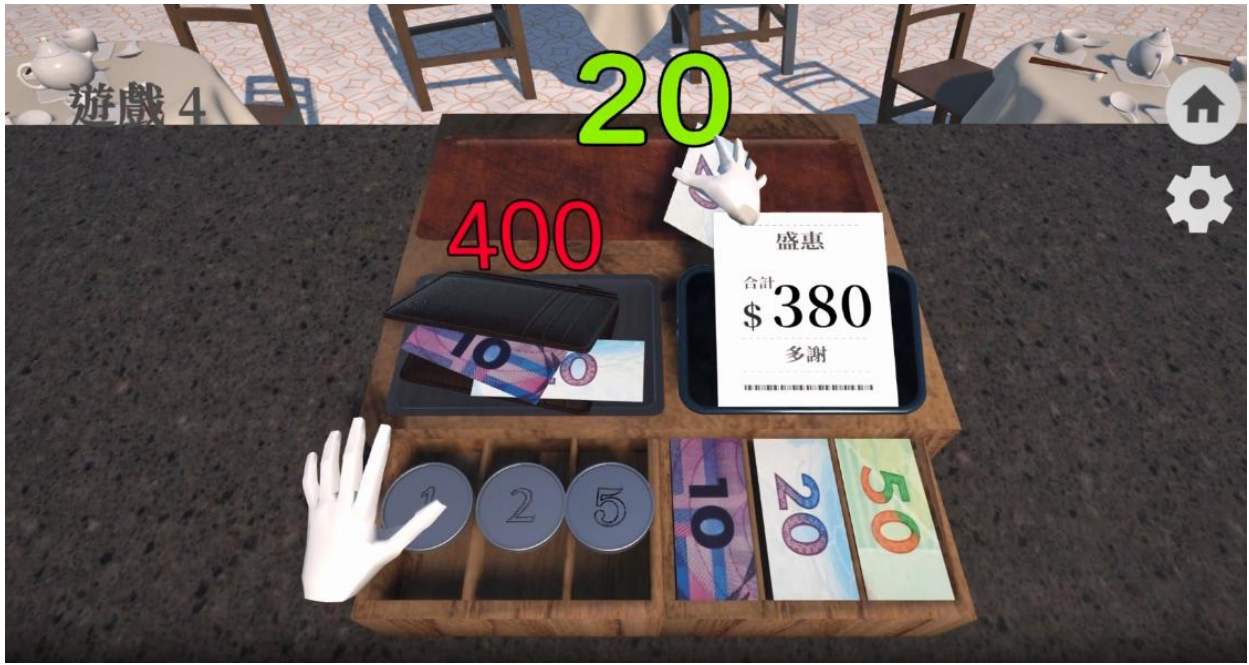

*Figure 5. Cashier game scene*

Within each module, the task flow follows a structured sequence guided by visual and auditory cues to facilitate comprehension and task completion. A Unity-based control panel allows researchers or therapists to adjust task difficulty and timing according to individual user capabilities. Performance data, including inaccuracy, omission, and task duration, are automatically recorded for quantitative analysis.

Overall, the SENSO™ architecture integrates immersive 3D design, gesture-based interaction, and IADL-inspired gameplay to create a motivating and therapeutic platform. This configuration contributes to cognitive stimulation, physical coordination, and everyday functional training for older adults with mild cognitive impairment.

**3.2 Study Design**

This study employed a single-session, within-subject usability and performance evaluation of the SENSO™ system among community-dwelling older adults. The primary objective was to assess system usability, learnability, perceived workload, and user satisfaction, and to characterize task performance across the three game modules.

Data collection took place in a classroom at the New Territories West Elderly Association Centre (NTWEAC), Hong Kong. The room was configured as a temporary experimental space, with standardized lighting, ambient noise control where feasible, and a consistent setup of the motion capture sensor, display, speakers, and olfactory device for all sessions.

Ethical approval was obtained from the Human Research Ethics Committee of UOW College Hong Kong (HREC012025). All participants provided written informed consent prior to enrollment.

### 3.3 Participants

A total of 50 community-dwelling older adults were initially recruited from NTWEAC through convenience sampling based on availability and interest. Inclusion criteria were:

- Age 60 years or above
- Community-dwelling status
- Sufficient physical function to perform sitting and standing tasks with upper-limb movements required by SENSO™
- Adequate corrected vision and hearing to perceive on-screen instructions and auditory cues
- Ability to understand and follow simple instructions in the study language

Individuals with a known diagnosis of dementia, acute medical conditions that could interfere with safe participation, or severe uncorrected sensory impairments were excluded. Although the SENSO™ system is intended for MCI rehabilitation, this initial evaluation focused on older adults without known dementia diagnoses to isolate usability and interaction issues from clinical symptomatology.

All participants completed a Hong Kong–Montreal Cognitive Assessment (HK-MoCA) screening at the beginning of the session to characterize cognitive status. Scores were later used descriptively in the analysis to contextualize performance and usability ratings.

Due to technical issues (e.g., sensor tracking failure) and occasional participant withdrawal during the session, complete datasets (all three game tasks plus questionnaires) were obtained from 41 participants, which served as the final sample for analysis.

Each participant attended a single, approximately 60-minute individual session conducted at NTWEAC. A trained researcher guided the session and provided safety monitoring and verbal support as needed. The procedure consisted of three main parts.

### Part I: Hong Kong-Montreal Cognitive Assessment (HK-MoCA)

Prior to game initiation, participants completed the Hong Kong-Montreal Cognitive Assessment (HK-MoCA) for rapid screening of mild cognitive impairment under the supervision of the researcher, which took approximately 15 minutes per participant. This validated instrument evaluates multiple domains including attention and concentration, executive functions, memory, language, visuospatial skills, conceptual thinking, and orientation.

Scores were recorded and stored in the SENSO™ system user profile for personalized game adaptation and longitudinal analysis.

**Part II: Game Play Session (Dim Sum, Steamer and Cashier)**

The researcher prepared the game setup for approximately 5 minutes prior to training initiation. Figures 6a and 6b depict the complete player configuration, including the motion capture sensor mounted on a clamp (Figure 6a), speaker positioned in front of the participant, and the olfactory device placed on the participant's shoulder (Figure 6b).

The researcher then guided participants through initial hand exercises to familiarize them with game controls, followed by the main modules (Dim Sum , Steamer and Cashier). The system collected and analyzed data throughout to evaluate cognitive performance, with a walkthrough tutorial for each game featuring animated hands, text, and audio guidance (Figure 7).

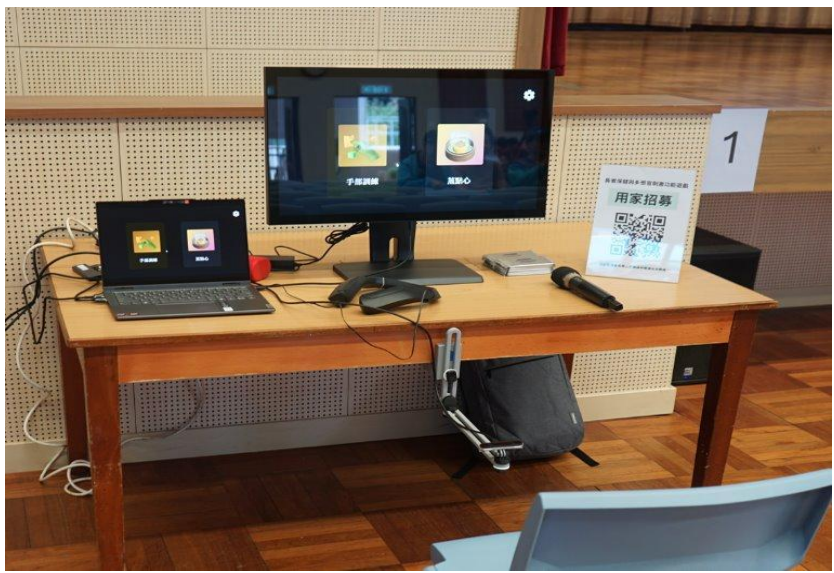

*Figure 6a: Setup of SENSO<sup>TM</sup>*

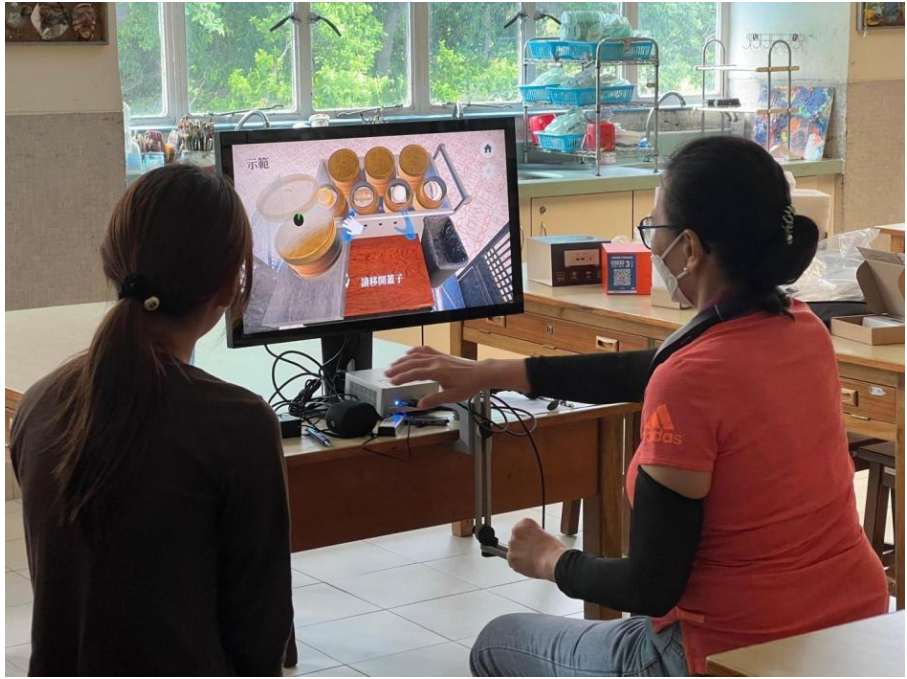

*Figure 6b: Setup with the player and researcher for verbal support.*

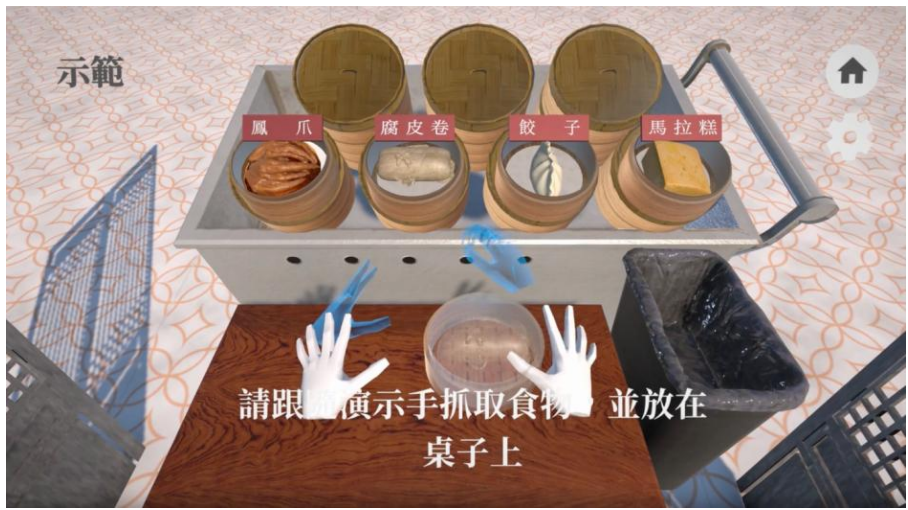

*Figure 7: Walkthrough tutorial for each game*

Participants then completed the three SENSO™ game tasks (Dim Sum, Steamer, and Cashier) in a fixed order. For each module, they first viewed a short on-screen explanation and then performed the task trials while the system automatically logged performance metrics (inaccuracy, omission, and task completion time). The researcher remained beside the participant, offering verbal clarification if needed, but refrained from providing hints that would alter task difficulty.

**Part III: Post-Experience Questionnaires (SUS, NASA-TLX, Custom Items)**

Following game play, 15-20 minutes required for the survey, participants completed the System Usability Scale (SUS) survey alongside custom questionnaires to quantify subjective usability and domain-specific feedback. SUS provided a standardized 10-item Likert-scale metric, with higher values indicating greater perceived ease of use for the SENSO™ multisensory interface.

Questionnaires targeted rehabilitation-specific aspects, including multisensory engagement, instructional clarity, and therapeutic motivation tailored to older adults. Responses enabled correlation with objective performance data for comprehensive usability validation and iterative system refinement.

## 3.6 Outcome Measures

### 3.6.1 Primary Outcome Measures

    A. **System Usability Scale (SUS)**

Usability evaluation of the SENSO™ system employs the standardized System Usability Scale (SUS) for quantitative assessment alongside custom questionnaires and a performance analysis form. SUS surveys capture subjective user perceptions on a 10-item Likert scale ranging from strongly disagree to strongly agree.

The SUS is a widely used, 10-item instrument with a 5-point Likert response format (1 = Strongly Disagree to 5 = Strongly Agree). It yields a global usability score from 0 to 100, capturing both usability and learnability aspects. Each item was scored according to standard procedures: for odd-numbered items, the score is (response – 1), and for even-numbered items, the score is (5 – response), with the sum multiplied by 2.5. Higher scores indicate better perceived usability.

Additionally, a questionnaire complemented to SUS is provided for the participants by probing domain-specific feedback such as ease of multisensory interaction and therapeutic engagement tailored for older adults in rehabilitation contexts.

### B. Performance Metrics

Performance on the three distinct SENSO™ games was evaluated using a framework proposed by Ranka and Chapparo (2011) for activity analysis. The system automatically recorded three key performance indicators for each task:

1. Inaccuracy: Percentage of incorrect actions relative to total required actions
2. Omission: Percentage of missed required actions relative to total required actions
3. Timing: Total task completion time seconds)

These indicators quantify accuracy, completeness, and efficiency of task execution within the SENSO™ environment.

**3.6.2 Secondary Outcome Measures**

*3.6.2.1 Demographics and Technology Experience.*

A custom questionnaire collected age, gender, education level, self-reported health status (including vision and hearing), and technology-related variables (e.g., gaming frequency, computer proficiency, prior experience with VR or motion capture).

*3.6.2.2 Pre-Experience Interest and Expectations*

Before gameplay, participants rated statements such as "If given the chance, I would participate in a gamified rehabilitation program" and "Offering functional games related to sports can attract more elderly individuals" on a 5-point Likert scale (1 = Strongly Disagree to 5 = Strongly Agree), capturing initial attitudes toward gamified rehabilitation.

*3.6.2.3 NASA Task Load Index (NASA-TLX)*

During the post-session phase, participants completed the NASA-TLX, rating six workload dimensions (Mental Demand, Physical Demand, Temporal Demand, Performance, Effort, Frustration) on a 7-point scale. Higher scores reflected greater perceived demand or, in the case of Performance, poorer perceived performance. These data provided insight into perceived task difficulty and suitability for the target population.

*3.6.2.4 Post-Experience Satisfaction and Perceived Usefulness.*

Additional Likert-scale items assessed overall satisfaction (e.g., "Overall, I am satisfied with the whole game experience"), enjoyment, perceived usefulness for rehabilitation, and motivation to use similar systems in the future.

### 3.6.3 Measurement Schedule

All measures were collected within a single testing session. HK-MoCA was administered at the beginning of the session (Part I). Performance metrics were recorded continuously and automatically during gameplay (Part II). SUS, NASA-TLX, and the custom questionnaires were completed immediately after gameplay (Part III).

## 3.7 Data Analysis

### 3.7.1 Statistical Software and Approach

All statistical analyses were conducted using IBM SPSS Statistics version 28 (IBM Corp., Armonk, NY). Data were analyzed at a significance level of $p = 0.05$. Given the relatively small sample size (N=41) and non-normal distribution of performance scores, non-parametric tests were employed where appropriate.

### 3.7.2 Demographic and Descriptive Statistics

Demographic characteristics (age group, gender, education level, technology experience) were summarized using frequencies and percentages for categorical variables and means with standard deviations (M ± SD) for continuous variables. HK-MoCA scores were reported overall and by age group to contextualize cognitive status; established interpretive guidelines (e.g., scores ≥ 26 indicative of generally intact cognition) were used descriptively rather than as strict inclusion/exclusion thresholds.

### 3.7.3 System Usability Scale (SUS) Analysis

SUS item responses were analyzed at two levels:

- Item-level analysis. Mean and standard deviation for each of the 10 SUS items were computed to identify specific strengths and weaknesses in system usability and learnability.
- Total SUS score. Global SUS scores (0–100) were calculated using the standard algorithm described above. Following Bangor et al. (2008), scores below 50 were interpreted as not acceptable, scores between 50 and 70 as marginal, and scores above 70 as acceptable to excellent.

### 3.7.4 Questionnaire Data Analysis

Data from the four-section questionnaire were analyzed as follows:

- Section A (Demographics and Background): Descriptive statistics only
- Section B (Pre-Experience Interest): Frequency distributions and percentages of Likert-scale responses (1-5 scale)
- Section C (NASA-TLX Task Load): Descriptive statistics (means, standard deviations) for the six subscales: Mental Demand, Physical Demand, Temporal Demand, Performance, Effort, and Frustration
- Section D (Post-Experience Satisfaction): Frequency distributions and percentages of Likert-scale responses

For the custom questionnaires and NASA-TLX:

- Demographic and background items were summarized descriptively.
- Pre-experience interest and post-experience satisfaction items were reported as frequency distributions and percentages across the 5-point Likert scale.
- NASA-TLX subscale scores were summarized using means and standard deviations to characterize perceived workload dimensions.

### 3.7.5 Performance Score Analysis

Performance on the three SENSO™ game tasks (Dim Sum, Steamer, Cashier) was evaluated using three performance indicators derived from the framework by Ranka and Chapparo (2011):

1. Inaccuracy: Percentage of incorrect actions
2. Omission: Percentage of missed required actions
3. Timing: Total time to complete task (seconds)

Kruskal-Wallis H test was used to conduct the statistical comparisons across age groups, which is a non-parametric one-way analysis of variance used to compare performance scores (inaccuracy, omission, timing) across the three age groups (60-69, 70-79, 80+ years). This test was justified by:

- Performance scores were not normally distributed (confirmed via Shapiro-Wilk test)
- Sample sizes were unequal and small, particularly in the 80+ group (n=2)
- The data represented ordinal-level measurements

### 3.7.6 Handling of Missing Data

Data completeness was high across all measures. The SENSO™ system automatically recorded performance data for all completed tasks, and questionnaires were administered under researcher supervision to minimize missing responses. Only participants with complete data for all three tasks and all questionnaires (N = 41) were included in the analyses. No imputation procedures were required.

### 3.7.7 Assumption Testing

Prior to conducting inferential tests, the following assumptions were verified:

1. Independence of observations: Participants were tested individually in separate sessions; ; no repeated measures across days were collected.
2. Measurement level: Performance scores represented at least ordinal-level data.
3. Distribution properties: Histograms, Q–Q plots, and Shapiro–Wilk tests were used to assess normality. Deviations from normality supported the use of non-parametric methods for performance comparisons.

1. **Results**

## 4.1 Demographics of Participants

The demographic data is presented in Table 1. The participants were divided into three age groups: 60-69 (n=22), 70-79 (n=17), and 80 or above (n=2). The total sample consisted of 41 participants, with 11 males (26.8%) and 30 females (73.2%). The 60-69 group had the highest proportion of females (81.8%), while the 80 or above group had an equal gender distribution.

Regarding educational attainment, the majority of participants across all groups had at least 7 years of education. The 60-69 group had the highest education level, with 72.7% of participants having 10 or more years of education. In contrast, the 80 or above group had the lowest education level, with one participant (50.0%) reporting only 0-3 years of education and another (50.0%) reporting 7-9 years of education. No participants in the oldest age group had more than 9 years of formal education.

Cognitive function, assessed by the Hong Kong Montreal Cognitive Assessment (HK-MoCA), revealed a mean score of 27.0 (SD = 2.6) across all participants, indicating generally intact global cognition. MoCA scores were similar in the 60–69 (M = 27.0) and 70–79 (M = 27.1) age groups, while the 80 or above group showed a lower average score (M = 24.5), though interpretation is limited due to the small sample size in this subgroup.

*Table 1. Demographic of all participants.*

|  | Age group (60-69) (n=22) | Age group (70-79) (n=17) | Age group (80 or above) (n=2) | All age group (n=41) |
|---|---|---|---|---|
| **Gender, n (%)** | | | | |
| Male | 4 (18.2) | 6 (35.3) | 1 (50.0) | 11 (26.8) |
| Female | 18 (81.8) | 11 (64.7) | 1 (50.0) | 30 (73.2) |
| **Education, n (%)** | | | | |
| 0 to 3 years | 0 (0) | 1 (5.9) | 1 (50.0) | 2 (4.9) |
| 4 to 6 years | 2 (9.1) | 3 (17.6) | 0 (0) | 5 (12.2) |
| 7 to 9 years | 3 (13.6) | 4 (23.5) | 1 (50.0) | 8 (19.5) |
| 10 to 12 years | 9 (40.9) | 7 (41.2) | 0 (0) | 16 (39.0) |
| Above 12 years | 7 (31.8) | 3 (17.6) | 0 (0) | 10 (24.4) |
| **MoCA** | 27.0 | 27.1 | 24.5 | 27.0 |

## 4.2 SUS Results

*Table 2. Result of system usability scale of all participants.*

| SUS questions | Average | Standard deviation |
|---|---|---|
| 1. I think that I would like to use this system frequently. | 3.34 | 1.22 |
| 2. I found the system unnecessarily complex. | 2.41 | 1.07 |
| 3. I think this system is very easy to use. | 3.56 | 0.92 |
| 4. I think that I would need the support of a technical person to be able to use the system. | 3.39 | 1.07 |
| 5. I found the various function of this system were well integrated. | 3.66 | 0.99 |
| 6. I thought there was too much inconsistency in this system. | 3.07 | 0.96 |
| 7. I would imagine that most people would learn to use this system very quickly. | 3.71 | 1.15 |
| 8. I found the system very cumbersome to use. | 2.68 | 1.11 |
| 9. I felt very confident to use the system. | 4.17 | 0.92 |
| 10. I needed to learn a lot of things before I could get going with the system. | 2.07 | 1.17 |

### 4.2.1 Identified findings

#### 4.2.1.1 User confidence in using the system

The results indicate a high level of user confidence when interacting with the system. Nearly half of the participants (46.3%, n = 19) strongly agreed with the statement "I felt very confident using the system," while an additional 34.1% (n = 14) agreed. In total, 80.4% of participants expressed confidence in operating the system.

The mean score of 4.17 (SD = 0.92) was substantially higher than the neutral midpoint of 3, suggesting that users, including older adults, felt comfortable and self-assured while engaging with SENSO™. This finding is particularly noteworthy given the motion-based control design, which appears to effectively support intuitive interaction across different levels of technological experience.

#### 4.2.1.2 Learnability

The results demonstrate strong user confidence in the system's ease of learning. Only a small proportion of participants—7.3% (n = 3) strongly agreed and 2.4% (n = 1) agreed—with the statement "I needed to learn a lot of things before I could get going with the system." In total, merely 9.7% of users perceived a steep learning requirement when using SENSO™.

The mean score of 2.07 (SD = 1.17), which falls well below the neutral midpoint of 3, further supports the system's high learnability. These findings suggest that users, including older adults, found SENSO™ intuitive and accessible, indicating its suitability for a wide user base.

#### 4.2.1.3 Critical Weaknesses

While the overall user feedback was positive, several aspects warrant further refinement, particularly considering that participants were first-time users of the SENSO™ system.

High dependency on support: Participants reported a moderate need for technical assistance when using the system (Q4: M = 3.39), which likely reflects an initial learning curve rather than a fundamental usability issue. With repeated exposure, such dependency is expected to decrease as users become more familiar with the interface.

Low desire for frequent use: User motivation to engage with the system regularly was moderate (Q1: M = 3.34), suggesting that incorporating progressive difficulty levels or reward-based elements may enhance long-term engagement.

Perceived inefficiency: The system was rated as somewhat cumbersome to operate (Q8: M = 2.68), which may stem from early adaptation challenges that could be mitigated through interface refinements and onboarding resources.

## 4.3 Questionnaire Results

### 4.3.1 Demographics and Background Information of Participants (Section A)

The participant cohort demonstrated a diverse range of gaming habits and levels of technological proficiency, with 26.8% engaging in gaming daily and 19.5% never playing games. Most identified as beginner to intermediate computer users, and while all used mobile devices, only 53.7% had prior VR experience and 29.3% had motion capture experience. Most participants indicated no visual, hearing, or physical impairments, confirming the cohort's suitability for the evaluation.

### 4.3.2 Pre-Experience Interest and Expectations (Section B)

Prior to engaging with the games, participants expressed a strong positive attitude toward the concept of game-based rehabilitation. The statement "If given the chance, I would participate in the gamified rehabilitation program" received predominantly favorable ratings, with 92.7% of participants selecting scores of 4 or 5. Similarly, the statement "Offering functional games related to sports can attract a larger number of elderly individuals" was also rated positively by 82.9% of participants.

These results suggest that participants held optimistic expectations regarding the applicability and appeal of gamified rehabilitation interventions for older adults.

### 4.3.3 Task Load and In-Experience Feedback (NASA-TLX) (Section C)

The NASA Task Load Index (NASA-TLX) results indicated a moderate to high perceived task load. Participants generally reported strong performance outcomes (76.6% rating success between 5 and 7), though the required effort was medium to high. Frustration levels remained low (70.7% selecting 1 or 2), suggesting the tasks were challenging yet accessible.

### 4.3.4 Post-Experience Satisfaction (Section D)

Following participation in the functional game sessions, participant feedback was overwhelmingly positive, reflecting a high level of satisfaction and acceptance of the system.

The statement "Overall, I am satisfied with the whole game experience" received very high ratings, with 85.4% of participants selecting scores of 4 or 5. Similarly, perceived usefulness increased after direct engagement, as demonstrated by responses to the statement "I like practicing through the gamified experience," which also received predominantly high ratings, with 73.2% of participants selecting 4 or 5.

Motivation for rehabilitation was also strongly affirmed, with 88.8% of participants rating 4 or 5 for the statement "Offering such functional games related to rehabilitation can help motivate more users."

Collectively, these findings suggest that the SENSO™ game experience effectively fostered user satisfaction, perceived usefulness, and motivational appeal, supporting its potential role as a positive and engaging tool in rehabilitation contexts.

Overall, the questionnaire findings reveal favorable user perceptions across all stages of engagement. Participants demonstrated diverse technological backgrounds yet expressed strong initial interest in gamified rehabilitation. During gameplay, task workload was rated as moderate to high but accompanied by low frustration levels, indicating effective task balance and accessibility. Post-experience feedback reflected high satisfaction, perceived usefulness, and motivation for continued use. Collectively, these results underscore the system's usability, acceptance, and potential to enhance engagement in technology-assisted rehabilitation among older adults.

## 4.4 Performance Analysis Results

The performance analysis was conducted following the framework proposed by Ranka and Chapparo (2011) using the SENSO™ scoring system across three game tasks: Dim Sum, Steamer, and Cashier.

Performance was evaluated using three indicators: Inaccuracy, Omission, and Timing, as the system does not record Repetition.

The Dim Sum Task involved grasping and placing dim sum items, the Steamer Task required sequential steaming and transferring actions, and the Cashier Task involved grasping change from a register. The results of the performance analysis are summarised in Table 3.

*Table 3. Performance analysis of SENSO$^{TM}$ game scores by age group.*

|  |  | Age group (60-69) (n=22) | Age group (70-79) (n=17) | Age group (80 or above) (n=2) | Significance between age groups |
| --- | --- | --- | --- | --- | --- |
| Dim Sum | Inaccuracy | 6.1 | 11.8 | 16.7 | 0.468 |
|  | Omission | 6.8 | 7.6 | 3.5 | 0.822 |
|  | Total time used (s) | 72.7 | 87.2 | 81.5 | 0.513 |
| Steamer | Inaccuracy | 4.9 | 8.3 | 0 | 0.261 |
|  | Omission | 17.4 | 17.8 | 62.5 | 0.003* |
|  | Total time used (s) | 186.7 | 213.2 | 478 | 0.022* |
| Cashier | Inaccuracy | 7.6 | 9.8 | 0 | 0.789 |
|  | Omission | 54.0 | 83.4 | 70 | 0.193 |
|  | Total time used (s) | 222.4 | 338.8 | 344.5 | 0.098 |

**4.4.1 Dim Sum Task**

Performance in the Dim Sum task remained stable across all age groups with low error and omission rates, and no statistically significant differences in completion time (p = 0.513), indicating the task was appropriately paced and accessible across the full age range.

#### 4.4.2 Steamer Task

The Steamer task demonstrated the most pronounced age-related differences, with the 80+ group showing significantly higher omission rates (62.5%, p = 0.003) and longer completion times (478 s, p = 0.022) compared to younger groups, reflecting its higher cognitive complexity and greater demands on working memory and attention.

#### 4.4.3 Cashier Task

Performance on the Cashier task displayed non-significant age-related trends toward slower completion times (from 222.4 s in the 60-69 group to 344.5 s in the 80+ group, p = 0.098) and relatively high omission rates across all groups (p = 0.193), suggesting multi-step activities posed occasional challenges, particularly for older participants.

Overall, the performance analysis revealed that participants were able to complete the SENSO™ game tasks with generally high accuracy and manageable task demands across age groups. The Dim Sum Task showed stable performance and minimal age-related effects, suggesting good accessibility for all participants. The Steamer Task demonstrated the most significant age-related differences, with the 80+ group exhibiting higher omission rates and longer completion times, indicating greater cognitive and attentional demands. The Cashier Task showed similar, though non-significant, trends toward slower performance and higher omission rates among older participants. Together, these findings highlight the SENSO™ system's ability to differentiate performance patterns across age groups while remaining broadly usable and engaging for older adults.

## Discussion and Future Directions

While the findings provide encouraging evidence of usability and task differentiation, they must be interpreted in light of sample and design constraints. The study cohort consisted predominantly of females and individuals with relatively high education in the younger age groups, which may have facilitated both technology adoption and cognitive task performance. In addition, all participants were community-dwelling and demonstrated generally intact cognition on the HK-MoCA, indicating that the sample represents a relatively healthy and motivated subset of older adults rather than the broader spectrum of aging and cognitive impairment. These factors limit generalizability to males, less educated or more frail individuals, and those with established MCI or dementia.

The single-session design, although appropriate for an initial usability and feasibility evaluation, precludes analysis of learning curves, long-term adherence, or sustained therapeutic effects. Performance metrics in this context reflect early interactions and may underestimate the extent to which users could adapt to the Steamer and Cashier tasks over repeated exposure. Conversely, high satisfaction and motivation ratings may partly reflect novelty effects that could diminish over time. Future research should therefore employ longitudinal designs in both healthy older adults and clinical MCI samples to examine how usability, workload, and performance evolve across multiple sessions and whether improvements translate into gains on standardized cognitive and functional outcomes.

## Conclusion

The SENSO™ system represents a successful integration of rehabilitation science and cutting-edge interaction technology. The findings confirm that healthy older adults not only accept multi-sensory VR interventions but can perform complex IADL-based tasks with high accuracy up until approximately 80 years of age. The "Steamer" task, with its high demands on executive sequencing and timing, serves as the most sensitive probe for age-related decline, providing a clear normative baseline for future MCI classification.

Moving forward, the integration of biomarkers, such as salivary DHEA or EEG-based engagement metrics, alongside game performance data will be essential to validate the system's predictive utility. By leveraging the "cognitive fingerprints" identified in this study, the SENSO™ platform is poised to become a vital tool in the non-pharmacological management of neurocognitive health, offering a bridge between controlled clinical assessments and the complex realities of independent daily living.

# Acknowledgements

We are grateful to New Territories West Academies Cluster, and staff for their support with this research.

# Declarations

### Availability of data and material

Access to relevant research data is available via the authors.

### Fundings

The project funding is partially supported by the 2nd City I&T Grand Challenge programme, organized by Hong Kong's Innovation and Technology Commission. UOW internal funding / Pathfinder Technology Limited funding

# Patents

VIRTUAL REALITY THERAPY SYSTEM (Hong Kong ref. no. 32024097392.7 and China ref. no. 202411353854.8)
COGNITIVE AND PHYSICAL DUAL-TASK COORDINATION TRAINING SYSTEM (Hong Kong ref. no. 22025109234.6 and China ref. no. 202510882255.3)

**Conflict of interest**

The author declares that they have no relevant financial or non-financial interests to disclose.

# References

Abd-Alrazaq, A., Alhuwail, D., Ahmed, A., & Househ, M. (2022). Effectiveness of serious games for improving executive functions among older adults with cognitive impairment: Systematic review and meta-analysis. JMIR Serious Games, 10(3), e36123. https://doi.org/10.2196/36123 https://doi.org/10.2196/36123

Aldardour, A., & Alnammaneh, S. (2025). The role of immersive virtual reality in geriatric rehabilitation. Rehabilitación, 59(3), 100930. https://doi.org/10.1016/j.rh.2024.100930

Anderson, N. D. (2019). State of the science on mild cognitive impairment (MCI). CNS Spectrums, 24(1), 78–87. https://doi.org/10.1017/s1092852918001347

Arshamian, A., Gerkin, R. C., Kruspe, N., Wnuk, E., Floyd, S., O'Meara, C., Croy, I., & Majid, A. (2022). The perception of odor pleasantness is shared across cultures. Current Biology, 32(9), 2061-2066.e4. https://doi.org/10.1016/j.cub.2022.03.047

Aïn, S. A., Poupon, D., Hétu, S., Mercier, N., Steffener, J., & Frasnelli, J. (2019). Smell training improves olfactory function and alters brain structure. NeuroImage, 189, 45–54. https://doi.org/10.1016/j.neuroimage.2019.01.008

Bishnoi, A., & Hernandez, M. E. (2020). Dual task walking costs in older adults with mild cognitive impairment: A systematic review and meta-analysis. Aging & Mental Health, 25(9), 1618–1629. https://doi.org/10.1080/13607863.2020.1802576

Bohil, C. J., Alicea, B., & Biocca, F. A. (2011). Virtual reality in neuroscience research and therapy. Nature Reviews Neuroscience, 12(12), 752–762. https://doi.org/10.1038/nrn3122

Brennan, P., Kaba, H., & Keverne, E. B. (1990). Olfactory recognition: A simple memory system. Science, 250(4985), 1223–1226. https://doi.org/10.1126/science.2147078

Brugada-Ramentol, V., Bozorgzadeh, A., & Jalali, H. (2022). Enhance VR: A multisensory approach to cognitive training and monitoring. Frontiers in Digital Health, 4. https://doi.org/10.3389/fdgth.2022.916052

Cotelli, M., Manenti, R., & Zanetti, O. (2012). Reminiscence therapy in dementia: A review. Maturitas, 72(3), 203–205. https://doi.org/10.1016/j.maturitas.2012.04.008

Gómez-Soria, I., Marin-Puyalto, J., Peralta-Marrupe, P., Latorre, E., & Calatayud, E. (2022). Effects of multi-component non-pharmacological interventions on cognition in participants with mild cognitive impairment: A systematic review and meta-analysis. Archives of Gerontology and Geriatrics, 103, 104751. https://doi.org/10.1016/j.archger.2022.104751

Hassandra, M., Galanis, E., Hatzigeorgiadis, A., Goudas, M., Mouzakidis, C., Karathanasi, E. M., Lianos, V., & Theodorakis, Y. (2021). A virtual reality app for physical and cognitive training of older people with mild cognitive impairment: Mixed methods feasibility study. JMIR Serious Games, 9(1), e24170. https://doi.org/10.2196/24170

Helbling, M., Grandjean, M., & Srinivasan, M. (2024). Effects of multisensory environment/stimulation therapy on adults with cognitive impairment and/or special needs: A systematic review and meta-analysis. Special Care in Dentistry, 44(2), 381–420. https://doi.org/10.1111/scd.12906

Honzíková, L., Dąbrowská, M., Skřinařová, I., Mullerová, K., Čecháčková, R., Augste, E., Salinger, V., & Štula, V. (2025). Immersive Virtual Reality as Computer-Assisted Cognitive–Motor Dual-Task Training in Patients with Parkinson's Disease. Medicina, 61(2), 248. https://doi.org/10.3390/medicina61020248

Khotbehsara, M., Soar, J., Lokuge, S., Khotbehsara, E., & Ip, W. K. (2025). The potential of virtual reality-based multisensory interventions in enhancing cognitive function in mild cognitive impairment: A systematic review. Journal of Clinical Medicine, 14(15), 5475.

Kim, O., Pang, Y., & Kim, J. (2019). The effectiveness of virtual reality for people with mild cognitive impairment or dementia: A meta-analysis. BMC Psychiatry, 19(1), 1–11. https://doi.org/10.1186/s12888-019-2180-x

Lam, W. W. T., Tang, Y. M., & Fong, K. N. K. (2023). A systematic review of the applications of markerless motion capture (MMC) technology for clinical measurement in rehabilitation. Journal of NeuroEngineering and Rehabilitation, 20(1), 1–25. https://doi.org/10.1186/s12984-023-01186-9

Lau, J. S. Y., & Agius, H. (2021). A framework and review of serious games for older adults with mild cognitive impairment. Multimedia Tools and Applications, 80(20), 30351–30394. https://doi.org/10.1007/s11042-021-11051-7

Lau, J. S., Tang, Y. M., Gao, G., Fong, K. N., & So, B. C. (2024). Development and usability testing of virtual reality (VR)-based reminiscence therapy for people with dementia. Information Systems Frontiers. https://doi.org/10.1007/s10796-024-10479-w

Lee, I. J., & Xin-Ting, P. (2024). Multisensory Virtual Reality Reminiscence Therapy: A Preliminary Study on the Initial Impact on Memory and Spatial Judgment Abilities in Older Adults. Human Factors in Virtual Environments and Game Design, 137.

Malloggi, E., Menicucci, D., Cesari, V., Frumento, S., Gemignani, A., & Bertoli, A. (2022). Lavender aromatherapy: A systematic review from essential oil quality and administration methods to cognitive enhancing effects. Applied Psychology: Health and Well-Being, 14(2), 663–690. https://doi.org/10.1111/aphw.12310

Manca, M., Paternò, F., Santoro, C., Zedda, E., Braschi, C., Franco, R., & Sale, A. (2021). The impact of serious games with humanoid robots on mild cognitive impairment older adults. International Journal of Human-Computer Studies, 145, 102509. https://doi.org/10.1016/j.ijhcs.2020.102509

Mancuso, V., Stramba-Badiale, C., Cavedoni, S., Pedroli, E., Cipresso, P., & Riva, G. (2020). Virtual reality meets non-invasive brain stimulation: Integrating two methods for cognitive rehabilitation of mild cognitive impairment. Frontiers in Neurology, 11. https://doi.org/10.3389/fneur.2020.566731

Manippa, V., Palmisano, A., Filardi, M., Nitsche, M. A., Rivolta, D., & Logroscino, G. (2022). An update on the use of gamma (multi)sensory stimulation for Alzheimer's disease treatment. Frontiers in Aging Neuroscience, 14, Article 1095081. https://doi.org/10.3389/fnagi.2022.1095081

Marin, C., Vilas, D., Langdon, C., Alobid, I., López-Chacón, M., Haehner, A., Hummel, T., & Mullol, J. (2018). Olfactory dysfunction in neurodegenerative diseases. Current Allergy and Asthma Reports, 18(8). https://doi.org/10.1007/s11882-018-0796-4

Moreno, A., Wall, K. J., Thangavelu, K., Craven, L., Ward, E., & Dissanayaka, N. N. (2019). A systematic review of the use of virtual reality and its effects on cognition in individuals with neurocognitive disorders. Alzheimer's & Dementia: Translational Research & Clinical Interventions, 5(1), 834–850. https://doi.org/10.1016/j.trci.2019.09.016

Muñoz, J., Mehrabi, S., Li, Y., Basharat, A., Middleton, L. E., Cao, S., Barnett-Cowan, M., & Boger, J. (2022). Immersive virtual reality exergames for persons living with dementia: User-centered design study as a multistakeholder team during the COVID-19 pandemic. JMIR Serious Games, 10(1), e29987. https://doi.org/10.2196/29987

Omon, K., Hara, M., & Ishikawa, H. (2019). Virtual reality-guided, dual-task, body trunk balance training in the sitting position improved walking ability without improving leg strength. Progress in Rehabilitation Medicine, 4, 20190011. https://doi.org/10.2490/prm.20190011

Optale, G., Urgesi, C., Busato, V., Marin, S., Piron, L., Priftis, K., Gamberini, L., Capodieci, S., & Bordin, A. (2010). Controlling memory impairment in elderly adults using virtual reality memory training: A randomized controlled pilot study. Neurorehabilitation and Neural Repair, 24(4), 348–357. https://doi.org/10.1177/1545968309353328

Rezaeyan, A., Asadi, S., Kamrava, S. K., Khoei, S., & Zare-Sadeghi, A. (2022). Reorganizing brain structure through olfactory training in post-traumatic smell impairment: An MRI study. Journal of Neuroradiology, 49(4), 333–342. https://doi.org/10.1016/j.neurad.2021.04.035

Riaz, W., Khan, Z. Y., Jawaid, A., & Shahid, S. (2021). Virtual reality (VR)-based environmental enrichment in older adults with mild cognitive impairment (MCI) and mild dementia. Brain Sciences, 11(8), 1103. https://doi.org/10.3390/brainsci11081103

Riley-Doucet, C. K., & Dunn, K. S. (2013). Using multisensory technology to create a therapeutic environment for people with dementia in an adult day center: A pilot study. Research in Gerontological Nursing, 6(4), 225–233. https://doi.org/10.3928/19404921-20130801-01

Rucco, R., Agosti, V., Jacini, F., Sorrentino, P., Varriale, P., De Stefano, M., Milan, G., Montella, P., & Sorrentino, G. (2017). Spatio-temporal and kinematic gait analysis in patients with frontotemporal dementia and Alzheimer's disease through 3D motion capture. Gait & Posture, 52, 312–317. https://doi.org/10.1016/j.gaitpost.2016.12.021

Schaumburg, M., Imtiaz, A., Zhou, R., Bernard, M., Wolbers, T., & Segen, V. (2025). Immersive virtual reality for older adults: Challenges and solutions in basic research and clinical applications. Ageing Research Reviews, 109, 102771. https://doi.org/10.1016/j.arr.2025.102771

Seyderhelm, A. J., & Blackmore, K. (2021). Systematic review of dynamic difficulty adaption for serious games: The importance of diverse approaches. SSRN Electronic Journal. https://doi.org/10.2139/ssrn.3982971

Slater, M. (2018). Immersion and the illusion of presence in virtual reality. British Journal of Psychology, 109(3), 431–433. https://doi.org/10.1111/bjop.12305

Sorokowski, P., Misiak, M., Roberts, S. C., Kowal, M., Butovskaya, M., Omar-Fauzee, M. S., Huanca, T., & Sorokowska, A. (2024). Is the perception of odour pleasantness shared across cultures and ecological conditions? Biology Letters, 20(6), 20240120. https://doi.org/10.1098/rsbl.2024.0120

Soudry, Y., Lemogne, C., Malinvaud, D., Consoli, S., & Bonfils, P. (2011). Olfactory system and emotion: Common substrates. European Annals of Otorhinolaryngology Head and Neck Diseases, 128(1), 18–23. https://doi.org/10.1016/j.anorl.2010.09.007

Sánchez, A., Millán-Calenti, J. C., Lorenzo-López, L., & Maseda, A. (2013). Multisensory stimulation for people with dementia. American Journal of Alzheimer's Disease & Other Dementias, 28(1), 7–14. https://doi.org/10.1177/1533317512466693

Tortora, C., Di Crosta, A., La Malva, P., Prete, G., Ceccato, I., Mammarella, N., Cacciari, C., & Palumbo, R. (2024). Virtual reality and cognitive rehabilitation for older adults with mild cognitive impairment: A systematic review. Ageing Research Reviews, 93, 102146. https://doi.org/10.1016/j.arr.2023.102146

Wei, X., Huang, C., Ding, X., Zhou, Z., Zhang, Y., Feng, X., Hou, Y., & Lü, J. (2025). Effect of virtual reality training on dual-task performance in older adults: A systematic review and meta-analysis. Journal of NeuroEngineering and Rehabilitation, 22(1), 141. https://doi.org/10.1186/s12984-025-01188-5

Wenk, N., Buetler, K. A., Penalver-Andres, J., Müri, R. M., & Marchal-Crespo, L. (2022). Naturalistic visualization of reaching movements using head-mounted displays improves movement quality compared to conventional computer screens and proves high usability. Journal of NeuroEngineering and Rehabilitation, 19(1), 137. https://doi.org/10.1186/s12984-022-01103-y

Woo, M., Kim, H. G., Kim, H., Cho, Y. S., & Kim, M. K. (2023). Olfactory enrichment in older adults during the night improves cognitive functioning: A pilot study. Scientific Reports, 13(1), 1162. https://doi.org/10.1038/s41598-023-28312-7

World Health Organization. (2023, March 15). Dementia. https://www.who.int/news-room/fact-sheets/detail/dementia

Yang, H., Luo, Y., Hu, Q., Tian, X., & Wen, H. (2021). Benefits in Alzheimer's disease of sensory and multisensory stimulation. Journal of Alzheimer's Disease, 82(2), 463–484. https://doi.org/10.3233/jad-201554

Yun, S. J., Kang, M., Yang, D., Choi, Y., Kim, H., Oh, B., & Seo, H. G. (2020). Cognitive training using fully immersive, enriched environment virtual reality for patients with mild cognitive impairment and mild dementia: Feasibility and usability study. JMIR Serious Games, 8(4), e18127. https://doi.org/10.2196/18127

Zhao, Y., Feng, H., Wu, X., Du, Y., Yang, X., Hu, M., Zhang, H., Deng, S., Zhang, J., & Zhao, Y. (2020). Effectiveness of exergaming in improving cognitive and physical function in people with mild cognitive impairment or dementia: Systematic review. JMIR Serious Games, 8(2), e16841. https://doi.org/10.2196/16841

Zhu, S., Sui, Y., Shen, Y., Zhu, Y., Ali, N., Guo, C., & Wang, T. (2021). Effects of virtual reality intervention on cognition and motor function in older adults with mild cognitive impairment or dementia: A systematic review and meta-analysis. Frontiers in Aging Neuroscience, 13, 586999. https://doi.org/10.3389/fnagi.2021.586999